# Glassy Dynamics in a heavy ion irradiated NbSe$_2$ crystal

S. Eley[1], K. Khilstrom[2], R. Fotovat[2], Z. L. Xiao[2], A. Chen[3], D. Chen[4], M. Leroux[2], U. Welp[2], W.K. Kwok[2], L. Civale[1]*

[1]Condensed Matter and Magnet Science, MPA, Los Alamos National Laboratory, Los Alamos, NM 87545, USA
[2]Materials Science Division, Argonne National Laboratory, Argonne, IL 60439 USA
[3]Center for Integrated Nanotechnology (CINT), MPA, Los Alamos National Laboratory, Los Alamos, NM 87545, USA
[4]Materials Science and Technology Division, MST-8, Los Alamos National Laboratory, Los Alamos, NM 87545, USA
*Correspondence to: lcivale@lanl.gov

## Abstract

Fascination with glassy states has persisted since Fisher introduced the vortex-glass as a new thermodynamic phase that is a true superconductor that lacks conventional long-range order. Though Fisher's original model considered point disorder, it was later predicted that columnar defects (CDs) could also induce glassiness — specifically, a Bose-glass phase. In YBa$_2$Cu$_3$O$_{7-x}$ (YBCO), glassy states can cause distinct behavior in the temperature ($T$) dependent rate of thermally activated vortex motion ($S$). The vortex-glass state produces a plateau in $S(T)$ whereas a Bose-glass can transition into a state hosting vortex excitations called double-kinks that can expand, creating a large peak in $S(T$). Although glass phases have been well-studied in YBCO, few studies exist of other materials containing CDs that could contribute to distinguishing universal behavior. Here, we report on the effectiveness of CDs tilted ~30° from the $c$-axis in reducing $S$ in a NbSe$_2$ crystal. The magnetization is 5 times higher and $S$ is minimized when the field is parallel to the defects versus aligned with the $c$-axis. We see signatures of glassiness in both field orientations, but do not observe a peak in $S(T)$ nor a plateau at values observed in YBCO. We discuss the possibility that competing disorder induces a field-orientation-driven transition from a Bose-glass to an anisotropic glass involving both point and columnar disorder.

## Introduction

Fisher's pivotal paper[1] on vortex-glass superconductivity in disordered bulk materials described the state as hosting decaying metastable currents. Prior to this, it was known that in type-II superconductors, metastable currents decay logarithmically over time due to the cumulative dissipation introduced by thermally activated jumps of vortices out of pinning sites (defects). This phenomenon is known as flux creep, and creep measurements can provide experimental access to critical exponents associated with the vortex-glass phase, hence are useful for identifying and characterizing glassiness[2]. In fact, the primary objective of Fisher's paper was to show that a sharp equilibrium phase transition exists between the normal state [at high $T$ and fields ($H$)] and the flux creep phase at low $T$ and $H$.

He argued that a novel thermodynamic phase, the vortex-glass, appears below the phase boundary $T_g(H)$. Subsequently, Nelson and Vinokur[3,4] found similarities between the vortex-glass phase and their proposed Bose-glass phase hosted by materials containing correlated disorder (twin and grain boundaries, columnar defects). However, the mechanisms leading to the vortex-glass and Bose-glass phases are distinct. In the former, point disorder encourages wandering and entanglement of flux lines whereas, in the latter, vortices localize on extended, correlated defects[3]. The two states can be distinguished through measurements in tilted magnetic fields[4].

Besides the ability to induce glassiness, interest in columnar defects is further motivated by their strong pinning capacity, associated with large pinning energies and subsequent enhancements in the critical current density ($J_c$). Pinning from CDs is directional; that is, at high enough fields, pinning is strongest, therefore $J_c$ is highest, when the field is parallel to the CDs[5]. Despite the strong pinning capacity of CDs, YBCO crystals containing parallel CDs are known to demonstrate extremely high creep rates under certain measurement conditions. At low fields and with increasing temperature, the system evolves from a Bose-glass state hosting half-loop excitations to a non-glassy state in which the half-loops expand, connect with adjacent CDs, and form double-kinks (see Fig. 1). These kinks are unpinned or weakly pinned, therefore can slide relatively unhindered, which allows for rapid transfer of the vortex line between CDs and produces a prominent peak in $S(T)$[6]. The peak is quite large —several times higher than the plateau[7] in $S(T)$ at ~0.02-0.04 observed in pristine YBCO crystals and associated with a vortex-glass state. Furthermore, when the field is misaligned with the CDs, various staircase structures[8] (see Fig. 1a) are known to form; a distinct signature of such structures has not yet been identified in creep measurements.

Many studies have characterized the effects of columnar defects on $J_c$ ($\theta_H$)[5,9–17], where $\theta_H$ is the angle of the applied field. Much less is known about the effect of field orientation on the creep rate ($S$) and, more generally, creep in materials besides YBCO that contain CDs. For example, it is unknown why the peak associated with rapid double-kink expansion in YBCO has not been observed in other materials[18–22]. In this study, we characterize the effect of temperature, magnetic field and field orientation on vortex dynamics in a NbSe$_2$ crystal containing parallel CDs tilted ~30° from the $c$-axis. First, we observe the expected peak in $J_c(\theta_H)$ when **H** is parallel to the CDs, and we find that this peak is indeed accompanied by a dip in $S(\theta_H)$. Second, we compare and characterize $S(T)$ and $S(H)$ when the field is parallel to the defects (**H** || CDs) versus the $c$-axis (**H** || c). Last, we find evidence of glassiness in both field orientations.

## Sample fabrication and measurements

Our experiments are carried out on two undoped 2$H$-NbSe$_2$ crystals that were grown using iodine vapor transport[23]. One crystal was irradiated with 1.4 GeV $^{208}$Pb$^{56+}$ ions at a dose of 1.45x10$^{11}$ ions/cm$^2$ corresponding to a matching field of 3T (average distance between CDs ~ 26 nm) at the Argonne Tandem Linear Accelerator System (ATLAS) while mounted with the crystallographic $c$-axis ~30° from the incident beam. The low magnification Transmission Electron Microscopy (TEM) image of the irradiated crystal shown in Fig. 2a indicates that the columnar amorphous tracks are continuous and almost perfectly parallel to each other, consistent with previous studies[24] and with the small splay expected for 1.4 GeV Pb ions. Fig. 2b is a high magnification TEM image showing an angle of ~29° between the radiation direction and the NbSe2 [002] direction. The average diameter of the tracks is about 4 to 6 nm.

We chose an angle of ~30° from rather than parallel to the *c*-axis to distinguish the effects of the CDs from those of mass anisotropy and intrinsic correlated defects (e.g., edge and screw dislocations) that are known to produce a peak in $J_c(\theta_H)$ for **H** || c in YBCO[13]. Similarly, for tilted CDs, the mere existence of asymmetry between $J_c(\theta_H)$ and $J_c(-\theta_H)$ can provide evidence of correlated pinning.

The dimensions of the irradiated and pristine crystals are ~0.8 mm x 0.7 mm x 20 μm (length $L \times$ width $W \times$ thickness $\delta$) and ~1.5 mm x 0.3 mm x 8.5 μm, respectively. 2*H*-NbSe2 is a layered transition metal dichalcogenide with an s-wave gap structure that has attracted intense interest[25] because it hosts a coexisting incommensurate charge density wave phase and superconductivity below $T_c$~7 K. Our primary motivation for studying NbSe2 is that it is a clean system (few defects in as-grown crystals) that has a low Ginzburg number (*Gi*). Assuming a coherence length $\xi_{ab} \approx 8$ nm, penetration depth[26] $\lambda_{ab} \approx 250$ nm, and upper critical field anisotropy of $\gamma = H_{c2}^{ab}/H_{c2}^c = \xi_{ab}/\xi_c$~3.3 (all at *T*=0), we estimate[27] $Gi = (\gamma^2/2)\left[(\mu_0 k_B T_c)/\left(4\pi B_c^2(0)\xi_{ab}^3(0)\right)\right]^2 \approx 5 \times 10^{-6}$, where $B_c = \Phi_0/[2\sqrt{2}\pi\lambda_{ab}\xi_{ab}]$ is the thermodynamic critical field. Superconductors with low *Gi* can attain significantly lower creep rates[27] than superconductors with high *Gi*, such as YBCO (*Gi*~10⁻²). This evokes the question of whether glassy states in low *Gi* materials manifest as a plateau at such a high $S$~0.02 − 0.04 and double-kink expansion creates a peak at high *S*. More generally, it motivates garnering a better understanding of the dynamics of various vortex excitations and glassiness in materials with low *Gi*.

Magnetization (*M*) measurements were collected using a Quantum Design SQUID magnetometer with a rotating sample mount, and both transverse and longitudinal pick-up coils to measure each component of the magnetic moment. By measuring *M* versus *T* at 2 Oe, we find that the critical temperature of the irradiated crystal is $T_c \approx 7$ K, similar to that in pristine crystals[25]. We extracted $J_c(T)$ from the magnetization data using the Bean Model [28,29], $J_c(T) = 20\Delta M/W[1 - W/(3L)]$, for **H** || c, where $\Delta M$ is the difference between the upper and lower branches of the *M(H)* curve. For the data collected when **H** || CDs, the tilted field orientation weakens the Lorentz force seen by some of the circulating currents, necessitating a modification of the Bean model[30,31]: $J_c(T) = 20\Delta M/W[1 - W\cos(\theta_H)/(3L)]$. To measure creep, we record successive measurements of *M* every 15 s at fixed fields and temperatures, capturing the decay in the magnetization ($M \propto J$, where *J* is the induced current) over time (*t*). We then calculate the creep rate $S[T, \mathbf{H}(\theta_H)] = |d\ln J/d\ln t|$. See Methods for more details.

## Results and Discussion

**Magnetization in different field orientations.** Figure 3 compares isothermal magnetic hysteresis loops, *M(H)*, at *T*=1.8 K for the pristine crystal for **H** || c ($\theta_H$=0°), and the irradiated sample for both **H** || c and for the field aligned with the defects (**H** || CDs, $\theta_H = \theta_{CD} = -31°$). The pristine crystal demonstrates dramatically lower magnetization than

the irradiated crystal. This suggests a weak pinning landscape and that the columnar defects in the irradiated crystal are overwhelmingly the predominant source of pinning.

For the irradiated crystal, the magnetization is roughly 5 times higher when the field is aligned with the CDs than with the *c*-axis. A large enhancement was anticipated and had been observed in previous studies, though the magnitude was less[5]. This improvement could be attributed to the higher energy used during irradiation (1.4 GeV Pb$^{56+}$ versus 300 MeV Au$^{26+}$ in Ref. [5]), which might create straighter, more continuous tracks.[32]

The dip at low fields $\mu_0 H <$ 0.6 T is caused by the out-of-plane pinning anisotropy. That is, pinning by extended defects along the *c*-axis (or, in our case, tilted 30° from) should produce a weak dip in *M*(*H*) at zero field, while pinning along the crystallographic *ab*-plane is expected to produce a peak[33]. At fields below self-field $H_{sf} \gg H$, vortex lines over a large region of the sample peripheries are quite curved. As the applied field is increased (approaching self-field), this region decreases as vortices straighten over a wider portion of the sample center. Columnar defects are more effective at pinning straight vortices. Hence, the initial increase in *M* with increasing *H* is caused by a combination of the heightened effectiveness of individual CDs in pinning less curved vortices and growing portions of the sample containing straight vortices. Predicted theoretically [33], the peak has been observed in irradiated YBCO[34] and Ba(Fe$_{0.93}$Co$_{0.07}$)$_2$As$_2$ crystals[35].

Additional *M*(*H*) loops were collected at *T*=4.5 K and at 20 different angles. Select curves are shown in Figs. 4a and 4b, capturing crossovers into different regimes. Note that the curves converge near zero field. This is because in the very dilute limit and for all field orientations, vortex lines will be oriented normal to the sample surface (aligned with the *c*-axis) to minimize their energy by shortening [5].

As the field tilts away from alignment with the CDs ($|\theta_H - \theta_{CD}| > \sim 6°$), the low-field peak progressively shifts to lower fields and eventually disappears. In particular, at $\theta_H$=-24°, *M*(*H*) decreases nearly linearly with decreasing *H*. Further rotation of the field away from the CDs ($\theta_H$<-40°, $\theta_H$>-19°) changes the *M*(*H*) behavior. *M* initially abruptly decays with increasing *H*, showing similar shape to *M* when **H**||c (Fig. 3). As the field is increasingly tilted ($\theta_H \geq$-2°), the *M*(*H*) curves exhibit a weak second magnetization peak (known as the fishtail effect) between 0.5 T and 1 T. This is most pronounced at $\theta_H$=33°, as highlighted in Fig. 4b. The fishtail effect has been observed in a wide variety of materials, including low-temperature superconductors, cuprates, MgB$_2$, and iron-based superconductors [36,37] and associated with an equally wide variety of effects, including elastic-to-plastic crossovers, vortex order-disorder phase transitions, and vortex lattice structural transitions [36]. In fact, a previous study[38] reported the appearance of a fishtail in a pristine NbSe$_2$ crystal when the applied field was tilted 30° from the *c*-axis and attributed it to a vortex order-disorder transition.

Extracted from the *M*(*H*) loops, the data is re-plotted as *M*($\theta_H$) at different fields in Fig. 5. The peak at $\theta_H$=$\theta_{CD}$ is clear at all fields and *M* rapidly decays at the slightest field misalignment with the defects, corresponding to a large reduction in $J_c$. For example, if we compare critical currents when the field is aligned with the CDs versus the c-axis, we find that $J_c$ is ~240 kA/cm$^2$ compared to ~48 kA/cm$^2$, respectively, at 0.6 T. Figure 6 shows such a comparison at 0.3 T over a broad temperature range, displaying an increase in $J_c$ by a factor of ~4 at 4.5 K and ~3 at 1.8 K. Note that the defects are effective even down to the lowest field of 0.2 T, where $J_c$ is only ~ 10% lower than at the maximum. This is consistent with all data in Fig. 5 being well above $H_{sf} \sim J_c\delta \leq 550$ Oe at this temperature. At most angles, lower fields produce higher $M$. However, for $\theta_H > 0°$, some low field curves cross, resulting in non-monotonic $M(H)$ that is consistent with the regime in which the fishtail is observed (Fig. 4b).

**Vortex creep when field is aligned with CDs versus *c*-axis.** To analyze vortex excitations and the potential for glassy dynamics, we measured the dependence of the creep rate on temperature and field orientation. First, we consider two creep models: the Anderson-Kim model and collective creep theory. A defect (or collection of defects) can immobilize a vortex segment (or a bundle of vortex lines) by reducing the vortex line energy by the pinning energy $U_P(T,H)$, which is the energy barrier that must be overcome for vortices to move. The Lorentz force induced by the persistent current $J$ then reduces $U_P$ to an activation barrier $U_{act}(T, H, J)$ and the vortex hopping rate is $\sim e^{-U_{act}/k_B T}$. The Anderson-Kim model[2], which neglects vortex elasticity and therefore does not predict glassy behavior, often accurately describes creep at low temperatures $T \ll T_c$. It assumes $U_{act}(J) \propto U_P |1 - J/J_c|$ for $J \lesssim J_c$. As $U_P$ is nearly temperature-independent at low $T$, $S$ is expected to increase linearly with increasing $T$, resulting in[2] $S(T) \approx k_B T / U_P$. At high temperatures, $S(T)$ steepens as $U_P(T)$ decreases.

Collective creep theory[2] predicts that the temperature dependence of the creep rate is

$$S = \frac{k_B T}{U_P + \mu k_B T \ln(t/t_0)}, \quad (1)$$

where $t_0$ the effective hopping attempt time and $C \equiv \ln(t/t_0) \sim 25 - 30$. Here $\mu > 0$ is the glassy exponent indicating the creep regime: $\mu$ = 1/7, 3/2 or 5/2, and 7/9 are predicted for creep of single vortices, small bundles (size less than the penetration depth $\lambda_{ab}$) and large bundles (size greater than $\lambda_{ab}$) of flux, respectively. At low temperatures $T \ll T_c$, $U_P \gg \mu k_B T \ln(t/t_0)$ such that $S(T) \approx k_B T/U_P$, coinciding with the Anderson-Kim prediction.

We now compare creep data for the irradiated crystal in two different field orientations: **H** || CDs and **H** || c. Note that our measurements are restricted to low fields because at high temperatures and fields, the magnetic signal is quite small when **H** ||c. Figure 7a shows the measured creep rate versus field orientation at 4.5 K and 0.5 T. Creep is clearly minimized when the field is aligned with the defects; $S$ is an order of magnitude smaller for **H** || CDs than for **H** || c. In fact, aligning the field with the defects suppresses creep at all fields and temperatures measured in our study, e.g., the comparison of $S(H)$ in both field orientations at 1.8 K shown in Fig. 7b.

Comparing creep data for the irradiated sample to the pristine crystal can only be performed at very low fields because the measurement signal produced by the pristine crystal at higher fields is at the limit of our measurement sensitivity. The temperature dependence of the creep rate in the pristine crystal and the irradiated crystal at 0.02 T is shown in Fig. 7c. For both field orientations, $S$ increases linearly with $T$ up to 5.5 K, qualitatively adhering to the Anderson-Kim description. Despite the very low applied field, the CDs are effective at lowering creep when **H** || CDs, but not when **H** || c, seen from a comparison to the data from the pristine sample.

Considering collective creep theory, if $U_P \ll C\mu k_B T$, $S(T)$ should plateau at $S \sim 1/C\mu$. Such a plateau is predicted in the case of glassiness, such that $S \sim 0.02 - 0.04$, equivalent to typical observations of plateaus in YBCO single crystals[7] and iron-based superconductors[31,39-44]. For our NbSe$_2$ crystal, Fig. 7d shows $S(T)$ at $\mu_0 H$=0.3 - 0.5 T for the two field orientations. In all cases, in Figs. 7c and 6d the creep rates are well below the usual collective creep plateau. The simplest interpretation is that $U_P$ is not negligible compared to $C\mu k_B T$ (see eq. 1), which is in agreement with the pinning energy estimates described below. Although, consistent with this scenario, most of the $S(T)$ curves in Figs. 7c and 6d are monotonically increasing, Fig. 7d also shows a broad temperature insensitive region in the 0.5 T data for **H** || c ($S \sim 0.003$) and a narrower one in the 0.3 T data for **H** || CDs ($S \sim 0.002$). However, interpretation of these data as indicative of a plateau at much lower than usual values would imply $C\mu \sim$ 300-500, requiring unphysically large values of either μ (10-17) or C (120-200), and consequently can be disregarded (note that typical $C$ and $\mu$ values give $C\mu$ < 75). Finally, quantum creep may be a significant component of our measured creep rates at these low temperatures, in which case, adding a temperature independent (and unfortunately unknown) contribution would imply an even smaller thermal creep contribution.

A plateau in $S(T)$ is the most apparent manifestation of glassy vortex dynamics. In its absence, we need a different approach to assess the nature of the vortex depinning excitations. Analysis of the current dependence of the effective activation energy $U^* \equiv T/S$ can provide direct experimental access to μ without the need for assumptions regarding $U_P$. According to collective creep theory[2], the activation barrier depends on the current as

$$U_{act}(J) = \frac{U_P}{\mu}\left[\left(\frac{J_{c0}}{J}\right)^\mu - 1\right], \quad (2)$$

where $J_{c0}$ is the temperature-dependent critical current in the absence of flux creep. Considering the Arrhenius hopping rate $\sim t_0^{-1} e^{-U_{act}(J)/k_B T}$ and equations (1) and (2), the effective pinning energy is

$$U^* \equiv \frac{T}{S} = U_P \times (J_{c0}/J)^\mu, \quad (3)$$

where μ > 0 for glassy creep and $\mu \to p < 0$ for plastic creep[45]. The exponent can thus easily be extracted from the slopes of $U^*$ vs $1/J$ on log-log plot. From Fig. 8, we see distinct elastic-to-plastic crossovers for all sets of data. At low $T$ the dynamics is clearly glassy at both field orientations, with μ~1. This is one of the main experimental findings of this study. As $T$ increases the dynamics turns plastic, with $p$ in agreement with the expectation for the motion of dislocations in the vortex lattice ($p$=-0.5) [46].

For **H** || CDs, glassy dynamics with μ~1 could be expected for a Bose-glass state characterized by half-loop formation. However, glassiness was unforeseen for **H** || c. In this configuration, we expected to see evidence of staircase structures (see Fig. 1), which form when the field is tilted away from the CDs by an amount greater than the lock-in angle ($\theta_L$), but less than the trapping angle ($\theta_t$). Yet in the simplest scenario staircases should be non-glassy, as finite length kinks easily slide along CDs. So, several possibilities should now be considered. Either $\theta_H = 0°$ is within the lock-in angle and half-loop excitations are

responsible for $\mu \sim 1$, or the dynamics of the staircase vortices is glassy, or this orientation is beyond $\theta_L$ and the CDs do not produce correlated pinning (so glassiness arises from standard random collective pinning).

A Bose-glass state formed when the field is aligned with CDs (and vortices are localized on these defects) will be robust to small changes in field orientation. That is, when the field tilted away from the CDs by an angle less than $\theta_L$, vortices will remain completely pinned by the CDs. This results in a plateau in $M(\theta_H)$ for $|\theta_H - \theta_{CD}| < \theta_L$ that has been observed in cuprates[47–50] and Co-doped BaFe$_2$As$_2$ [14]. Though our data is too coarse to determine if there is a lock-in effect and identify $\theta_L$, we see from Fig. 5 that the magnetization is greatly reduced at $\theta_H = 0°$ versus $\theta_H = \theta_{CD}$. So, $\theta_H = 0°$ is clearly well beyond the lock-in angle. On the other hand, the asymmetry of $M(\theta_H)$ around $\theta_H = 0°$, which can only arise from the tilted CDs, suggests that staircases are present at this orientation[47].

Having eliminated half-loops and random collective pinning as the cause of $\mu \sim 1$ at **H** || c, we consider the possibility of a vortex-glass state or an anisotropic glass involving both columnar and point disorder, as predicted in Ref. [ 4]. Segments of a single vortex line could be alternatingly pinned by adjacent CDs and interstitial point defects. As the current and thermal energy act on the vortex, the segments pinned by point defects might wander/entangle (instead of sliding like kinks). Alternatively, interactions among weakly pinned kinks may create "kink bundles" that, by analogy with the 3D vortex bundles, should exhibit glassy collective creep with μ~1. In either case, if the phase for **H** || CDs is indeed a Bose-glass then the system experiences a field-orientation-driven transition from a Bose-glass (**H** || CDs) to an anisotropic glass (**H** || c). However, an alternative interpretation is that, even for **H** nominally parallel to the CDs, due to a small misalignment the system is outside the lock-in regime and we are not observing a Bose glass phase. This could occur as a consequence of a very small $\theta_L$, as expected from theoretical estimates and confirmed by the sharpness of the cusp at $\theta_{CD}$ in Fig. 5. If that is the case, then we are observing staircases in both configurations and the differences in $J_c$ and $S$ arise from the much larger number of kinks for **H** || c. Additional studies with an angular resolution finer than $\theta_L$ would be needed to elucidate this point.

**Pinning energies.** The effectiveness of CDs is typically assessed by evaluating the measured pinning energies, which can be calculated from the creep data. The scale of the pinning energy in a superconductor[51] is approximately the condensation energy $U_{P1} \sim (B_c^2/2\mu_0)V$ within a coherence volume $V \sim V_c = (4\pi/3)\xi_{ab}^3/\gamma$. For NbSe$_2$, we calculate that $U_{P1} \sim 160 - 300$ K within our measurement $T$ range. From the Fig. 7a inset, we see that the effective activation energies $U^*$ extracted from our creep measurements plummets from being considerably greater than to comparable to $U_{P1}$ as the field rotates from **H** || CDs to **H** ||c. This is because pinning energies larger than $U_{P1}$ can be achieved through individual strong pinning by defects larger than $V_c$, as is the case for our CDs.

Columnar defects are most effective at pinning vortices of smaller size $\xi_{ab} \leq 2R$ (where $R$ is the CD radius)[12]. This is not easily achieved in low-$T_c$ superconductors, which tend to have large coherence lengths. When $R < \xi_{ab}\sqrt{2}$ (as is the case for our sample), the vortex binding energy per unit length is $U_0 = \varepsilon_0 (R/2\xi_{ab})^2$ where $\varepsilon_0 = \Phi_0^2/(4\pi\mu_0\lambda_{ab}^2)$ is the line energy. A vortex pinned to an isolated CD may depin when the half-loop length is $\ell_{hl} \sim [U_0\varepsilon_\ell/\varepsilon_0^2]^{1/2}(\xi_{ab}/\gamma)(J_0/J_c)$ (the half-loop nucleus reaches a critical radius) and the associated pinning energy is $U_{h\ell} \sim U_0 \ell_{hl}$. Here, $\varepsilon_\ell = \varepsilon_0 \ln(\xi_{ab}/\lambda_{ab})$ and $J_0 = \Phi_0/(3^{3/2}\pi\mu_0\lambda_{ab}^2\xi_{ab})$ is the depairing current[2]. Note that the transverse size of the half-loop depends on competition between the elastic energy $\varepsilon_l U_{hl}/l_{hl}$ and pinning energy $U_{h\ell}$, and that the critical size occurs when the Lorentz force $J\Phi_0 l_{hl} U_{hl}/c$ matches the elastic energy[2]. A system containing half-loops therefore exhibits a glassy response because the half-loop energy barrier increases with decreasing current. For our NbSe$_2$ sample when **H** || CDs, we calculate $U_{h\ell} \sim 600$ K using[52,53] $\xi_{ab}(4.5 \text{ K}) = [\Phi_0/2\pi H_{c2}(4.5 \text{ K})]^{1/2} \sim 12.5$ nm, $\lambda_{ab}(4.5 \text{ K}) \sim 200$ nm, $R \sim 2.5$ nm and our measured $J_c(4.5 \text{ K}, 0.5 \text{ T}) \sim 230$ kA/cm$^2$. Our experimental $U^*(4.5 \text{ K}, 0.5 \text{ T}) = T/S \sim 3500$ K is higher than this value. However, Differences could be accounted for by uncertainties in some of the parameters. In particular, the irradiation induced tracks may depress the superconducting order parameter over a distance larger than the diameter observed by TEM due to, e.g. lattice strain. For example, $R \sim 5.7$ nm yields $U_{h\ell} \sim 3500$ K).

## Conclusions

In conclusion, we have demonstrated that in a NbSe$_2$ crystal with tilted columnar defects, the critical current is maximized and creep is concomitantly minimized when the field is aligned with the defects ($T$=4.5 K, $\mu_0 H$=0.5 T). This result was not necessarily intuitive, as the rapid expansion of double-kinks can promote fast creep when **H** || CDs in YBCO (at low temperatures and fields $B < B_\Phi$). We also found that **H** || CDs preferentially produced lower creep rates than **H** || c over our entire measurement range, and that both field orientations resulted in glassy behavior. Many open questions remain, including why a distinct, large peak in $S(T)$ resulting from double-kink expansion has only been observed in YBCO. These results motivate further studies of creep rates associated with the Bose-glass state in other low $Gi$ materials containing columnar defects.

## Author Contributions
S.E. took the measurements, performed the data analysis, and wrote the paper. L.C. designed the experiment, and assisted in data analysis and manuscript preparation. R.F. and Z.L.X. grew the sample. K.K. assisted in the measurements and preparing and irradiating the crystals. M.L., U.W., and W.K. prepared and irradiated the sample. A.C. and D.C. obtained the TEM images. S.E., L.C., K.K., M.L., U.W., and W.K. discussed the results, implications, and commented on the manuscript.

## Acknowledgements
This work was funded by the U.S. DOE, Office of Basic Energy Sciences, Materials Sciences and Engineering Division. The work of A.C. and D.C. was supported by the NNSA's Laboratory Directed Research and Development (LDRD) Program and was performed, in part, at the Center for Integrated Nanotechnologies, an Office of Science User Facility operated for the U.S. Department of Energy (DOE) Office of Science.

## Methods

**TEM images.** The TEM specimen of the irradiated NbSe2 crystal was fabricated in a focused ion beam and the microstructure was characterized by using FEI Tecnai F30 Transmission electron microscopy (TEM, 300 kV).

**Magnetization Measurements.** Magnetization measurements were collected using a Quantum Design SQUID magnetometer with a rotating sample mount, and transverse and longitudinal pick-up coils to measure each component of the magnetic moment, $m_t$ and $m_l$, respectively. The angle of the field was verified by calculating $\theta_H = \tan^{-1}(m_t/m_l)$, the total moment $m = m_l \cos\theta_H$, and the magnetization $M = m/\delta LW$ (where $\delta$ μm is the thickness, $W$ is the width, and $L$ mm is the length). Creep data were taken using standard methods [7]. Firstly, the field was swept high enough ($\Delta H > 4H^*$) that the sample was fully penetrated with magnetic flux and in the critical state. Then, successive measurements of $M$ were recorded every 15 s, capturing the decay in the magnetization ($M \propto J$) over time ($t$). Last, the time was adjusted to account for the difference between the initial application of the field and the first measurement and $S = |d\ln M/d\ln t|$ is calculated from the slope of a linear fit to $\ln M$-$\ln t$. $T_c$ was determined from the temperature-dependent magnetization at $H$ = 2 Oe.

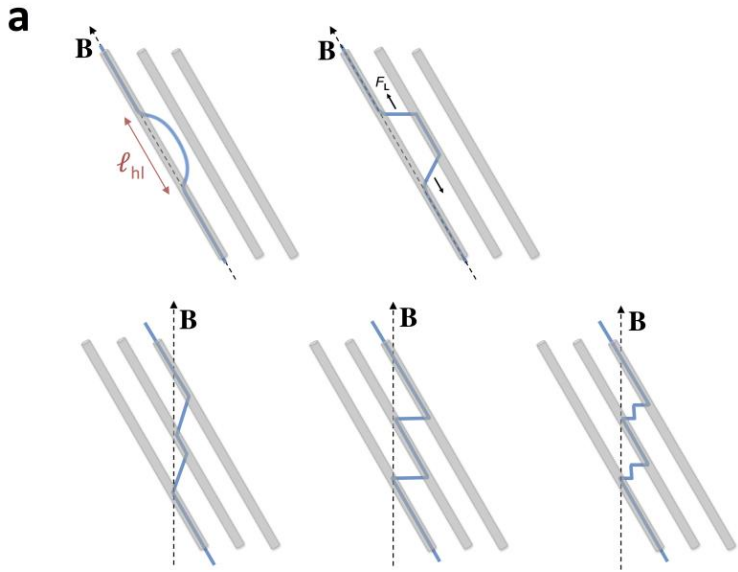

**Figure 1: Vortex structures involving columnar defects. a,** Illustration of possible vortex structures (blue lines) in samples with columnar defects (grey tubes). The dotted black lines indicate the direction of the applied field (**B**) and $\ell_{hl}$ labels the half-loop length. The upper panel shows half-loop (left) and double-kink (right) excitations. The lower panel illustrates possible staircase structures that form as a vortex line minimizes its energy by shortening and/or pinning to the *ab*-plane.

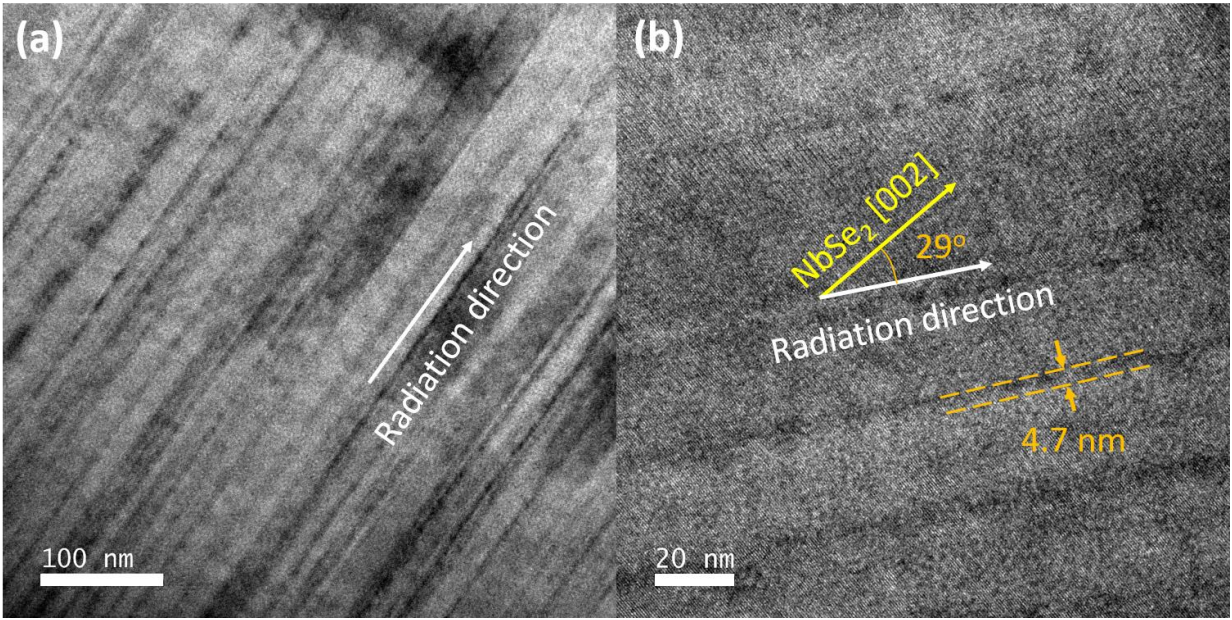

**Figure 2. a,** A low magnification TEM image of the irradiated NbSe₂ crystal showing continuous and parallel irradiation tracks. **b,** A high magnification TEM image showing an angle of ~28° between the radiation direction and the NbSe2 [002] direction. The average size of amorphous tracks is ~ 4.7 nm.

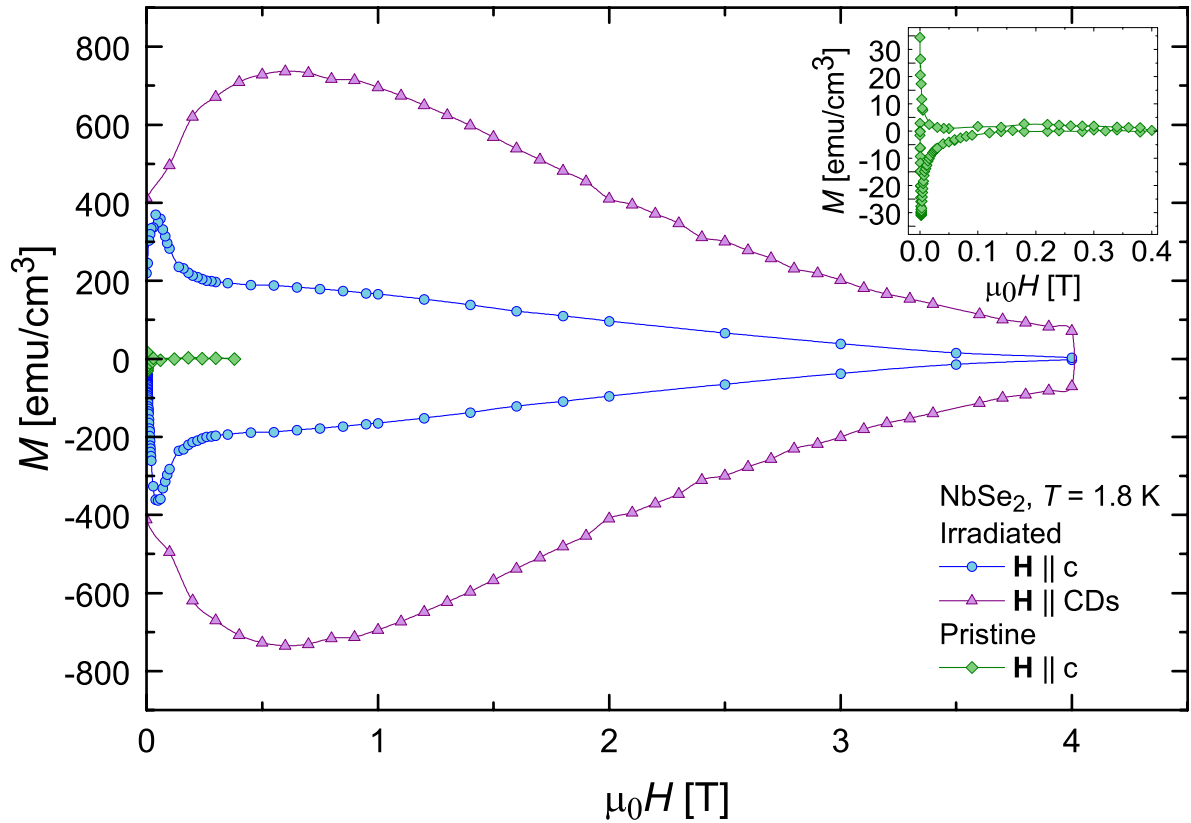

**Figure 3**: **Magnetic hysteresis loops when H is aligned with the columnar defects versus the *c*-axis.** Comparison of field dependent magnetization, $M(H)$, at $T$=1.8 K for the irradiated NbSe$_2$ crystal for two different field orientations (**H** parallel to *c*-axis versus **H** parallel to columnar defects) and the pristine NbSe$_2$ crystal for **H** parallel to the *c*-axis. **(inset)** Magnification of $M(H)$ loop for the pristine crystal.

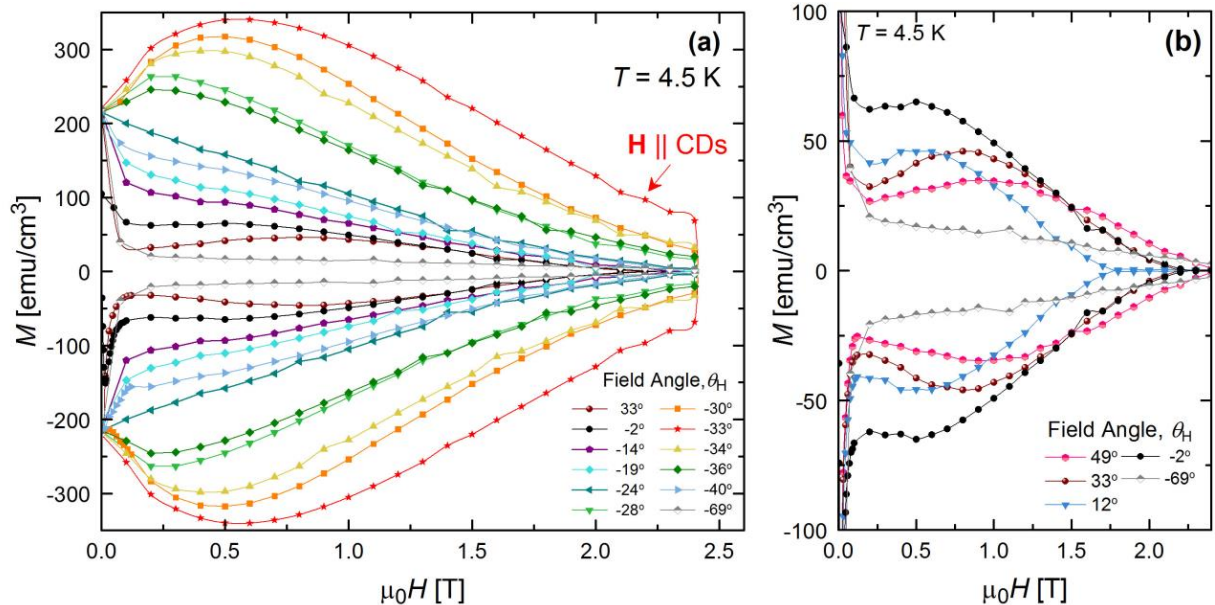

**Figure 4: Magnetic hysteresis loops for different field orientations**. **a,** Comparison of field dependent magnetization loops, $M(H)$, at $T$=4.5 K when the field is applied at several different angles ($\theta_H$). **b,** Select data from **a** highlighting the appearance of weak second magnetization peak when field is tilted far from defects ($\theta_H \geq -2°$).

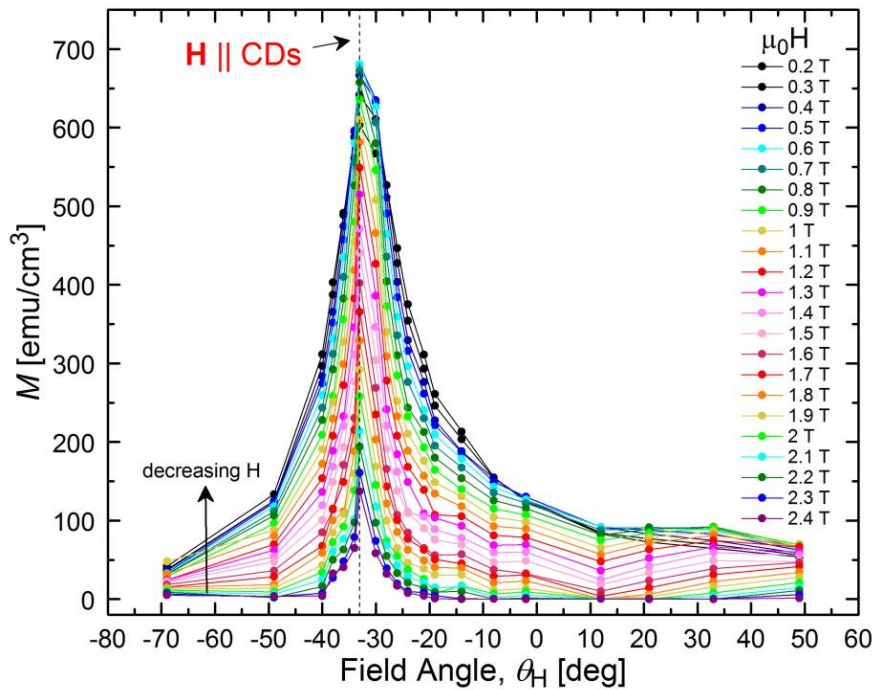

**Figure 5. Large peak in magnetization when applied field is aligned with columnar defects.** Magnetization ($M$) versus magnetic field orientation ($\theta_H$) in NbSe$_2$ crystal containing columnar defects tilted ∼-30° from the crystallographic $c$-axis. Data is shown for $T$=4.5 K and multiple values of the magnetic field.

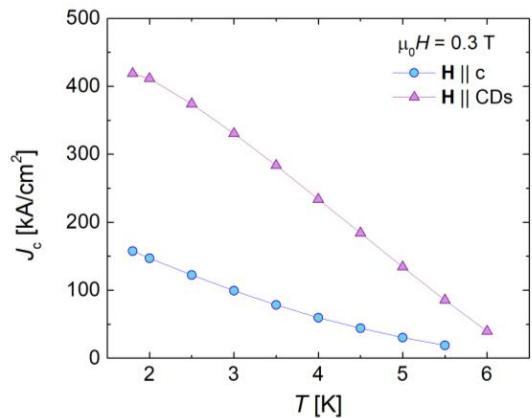

**Figure 6. Temperature dependence of the critical current.** Critical current ($J_c$) versus temperature for the irradiated crystal for both **H** || c and **H** || CDs in a field of 0.3 T.

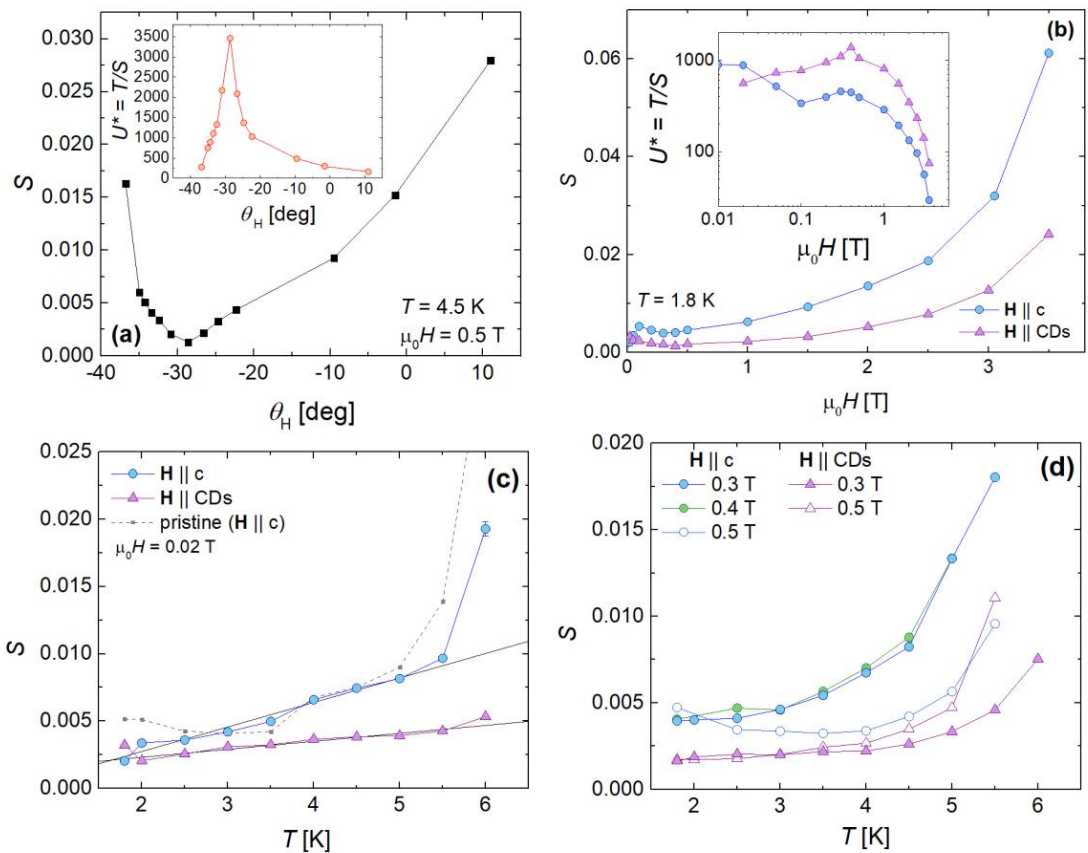

**Figure 7. Creep in irradiated NbSe₂ crystal. a**, Creep rate ($S$) versus magnetic field orientation ($\theta_H$) in the irradiated NbSe₂ crystal and **(inset)** the extracted effective pinning energy $U^*$. **b**, Comparison of field-dependent $S$ and **c**, **d**, temperature-dependent $S$ when the field is aligned with the columnar defects versus the c-axis. The solid black lines are linear fits to the low temperature data. The dotted grey curve in **c** is data for the pristine sample when the field is aligned with the c-axis.

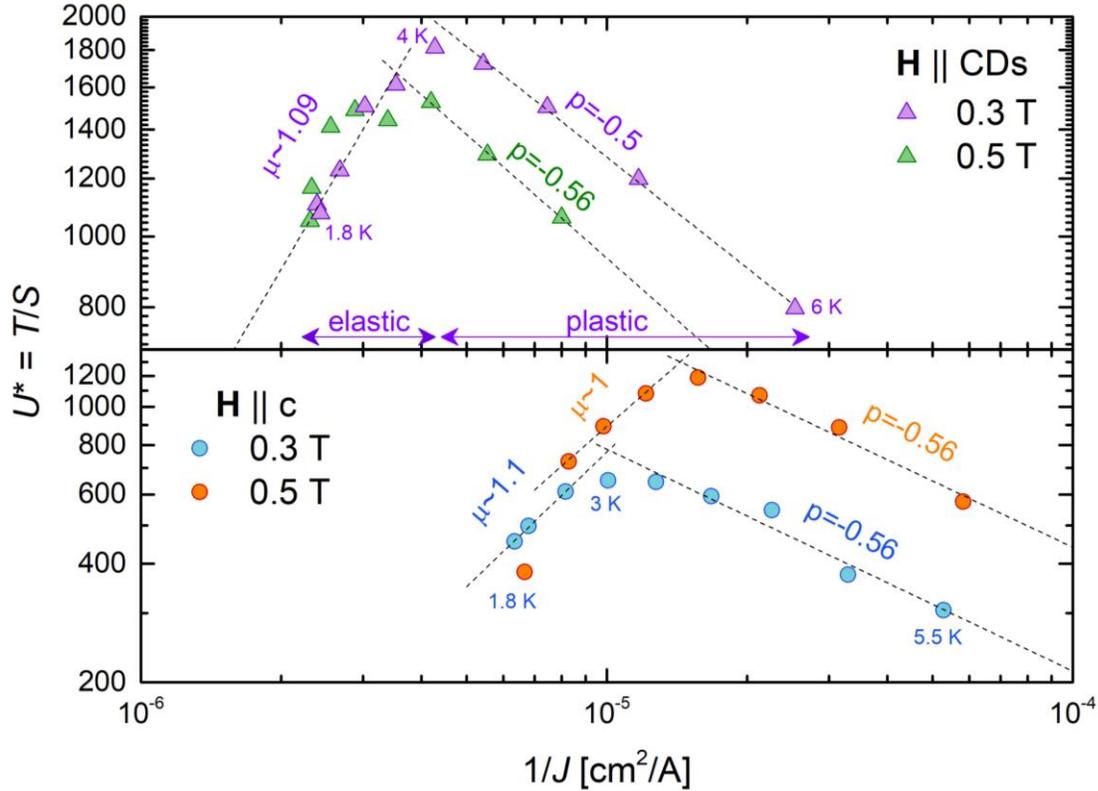

**Figure 8. Elastic to plastic crossovers.** Dependence of the effective energy barrier $U^*$ on the inverse current density when the field is aligned with the columnar defects (upper panel) versus the *c*-axis (lower panel). The extracted exponents μ and p are displayed in the plot, where $\mu = 1$ is expected for glassy behavior and p=-0.5 for plastic flow.